%% file: main.tex
\documentclass[conference]{IEEEtran}
\IEEEoverridecommandlockouts

\usepackage{amsmath,amssymb,amsthm,amsfonts}
\usepackage{algorithm} 
\usepackage{algpseudocode} 
\usepackage{graphicx}
\usepackage{textcomp}
\usepackage{xcolor}
\def\BibTeX{{\rm B\kern-.05em{\sc i\kern-.025em b}\kern-.08em
    T\kern-.1667em\lower.7ex\hbox{E}\kern-.125emX}}
    
\usepackage{tabularx}
\usepackage{makecell}
\usepackage{float}
\usepackage{booktabs} 
\usepackage{siunitx}
\usepackage{listings}
\usepackage[
    n,
    advantage,
    operators,
    sets,
    adversary,
    landau,
    probability,
    notions,    
    logic,
    ff,
    mm,
    primitives,
    events,
    complexity,
    oracles,
    asymptotics,
    keys
]{cryptocode}
\usepackage{comment}
\usepackage[colorinlistoftodos]{todonotes}
\usepackage{tikz}
\usepackage{pgfplots}
\pgfplotsset{compat=1.16}
\usepgfplotslibrary{statistics}
\usepgfplotslibrary{groupplots}
\usetikzlibrary{patterns}
\usepackage[framemethod=tikz]{mdframed} 
\usepackage{xurl}
\usepackage[hidelinks]{hyperref}
\hypersetup{bookmarks=false}

\input{ise}

\begin{document}
\include{style/defines}
\include{style/code}

\renewcommand\theadalign{bc}
\renewcommand\theadfont{\bfseries}
 
\theoremstyle{plain}
\newtheorem{thm}{Theorem}
\newtheorem{cor}[thm]{Corollary}
\theoremstyle{definition}
\newtheorem{definition}{Definition}

\createprocedureblock{procb}{center,boxed}{}{}{linenumbering}

\sisetup{
round-mode = figures,
round-precision = 3
}

\title{
Privacy-Preserving On-chain Permissioning for KYC-Compliant Decentralized Applications
}

\author{\IEEEauthorblockN{Fabian Piper, Karl Wolf, Jonathan Heiss}
\IEEEauthorblockA{\textit{Information Systems Engineering} \\
\textit{TU Berlin}\\
Berlin, Germany \\
{\{fpi,kw,jh\}}@ise.tu-berlin.de}
}
\maketitle

\input{contents4/00_Abstract}


\input{contents4/01_Introduction}
\input{contents4/02_Preliminaries}
\input{contents4/03_Model}
\input{contents4/04_System_Design}
\input{contents4/05_Practical_Realization}
\input{contents4/06_Evaluation}
\input{contents4/07_Discussion}
\input{contents4/08_Conclusion}

\section*{Acknowledgment}
Funded by the European Union (TEADAL, 101070186). Views and opinions expressed are, however, those of the author(s) only and do not necessarily reflect those of the European Union. 
Neither the European Union nor the granting authority can be held responsible for them.

\bibliographystyle{IEEEtran}
\bibliography{bibliography}

\newpage





\end{document}

%% file: ise.tex
\newcommand{\step}[1]{\tikz[baseline=(char.base)]{
            \node[shape=circle,draw,inner sep=1pt] (char) {#1};}}

\usepackage{comment}

%% file: style/defines.tex
\newcommand{\hdv}{\pcadvstyle{H}}
\newcommand{\vdv}{\pcadvstyle{V}}
\newcommand{\odv}{\pcadvstyle{O}}
\newcommand{\idv}{\pcadvstyle{I}}

\newcommand*\circled[1]{\tikz[baseline=(char.base)]{
            \node[shape=circle,draw,fill=black,text=white,inner sep=2pt] (char) {#1};}}

\newcommand{\numusd}[1]{\$\num[round-precision=2,round-mode=places,round-integer-to-decimal]{#1}}

%% file: style/code.tex
\definecolor{eclipseStrings}{RGB}{42,0.0,255}
\definecolor{eclipseKeywords}{RGB}{127,0,85}


\lstdefinelanguage{json}{
    basicstyle=\normalfont\ttfamily,
    commentstyle=\color{eclipseStrings}, 
    stringstyle=\color{eclipseKeywords}, 
    numbers=left,
    numberstyle=\scriptsize,
    stepnumber=1,
    numbersep=8pt,
    showstringspaces=false,
    breaklines=true,
    frame=lines,
    string=[s]{"}{"},
    comment=[l]{:\ "},
    morecomment=[l]{:"},
    literate=
        *{0}{{{\color{numb}0}}}{1}
         {1}{{{\color{numb}1}}}{1}
         {2}{{{\color{numb}2}}}{1}
         {3}{{{\color{numb}3}}}{1}
         {4}{{{\color{numb}4}}}{1}
         {5}{{{\color{numb}5}}}{1}
         {6}{{{\color{numb}6}}}{1}
         {7}{{{\color{numb}7}}}{1}
         {8}{{{\color{numb}8}}}{1}
         {9}{{{\color{numb}9}}}{1}
}

\lstset{
  escapeinside={(*}{*)},
  language=json,
  showstringspaces=false,
  extendedchars=true,
  basicstyle=\footnotesize\ttfamily,
  commentstyle=\slshape,
  %
  stringstyle=\ttfamily,
  breaklines=true,
  breakatwhitespace=true,
  %
  columns=flexible,
  numbers=left,
  numberstyle=\tiny,
  basewidth=.5em,
  xleftmargin=.5cm,
  %
  %
  %
  captionpos=b
}

\lstset{literate=
  {á}{{\'a}}1 {é}{{\'e}}1 {í}{{\'i}}1 {ó}{{\'o}}1 {ú}{{\'u}}1
  {Á}{{\'A}}1 {É}{{\'E}}1 {Í}{{\'I}}1 {Ó}{{\'O}}1 {Ú}{{\'U}}1
  {à}{{\`a}}1 {è}{{\`e}}1 {ì}{{\`i}}1 {ò}{{\`o}}1 {ù}{{\`u}}1
  {À}{{\`A}}1 {È}{{\'E}}1 {Ì}{{\`I}}1 {Ò}{{\`O}}1 {Ù}{{\`U}}1
  {ä}{{\"a}}1 {ë}{{\"e}}1 {ï}{{\"i}}1 {ö}{{\"o}}1 {ü}{{\"u}}1
  {Ä}{{\"A}}1 {Ë}{{\"E}}1 {Ï}{{\"I}}1 {Ö}{{\"O}}1 {Ü}{{\"U}}1
  {â}{{\^a}}1 {ê}{{\^e}}1 {î}{{\^i}}1 {ô}{{\^o}}1 {û}{{\^u}}1
  {Â}{{\^A}}1 {Ê}{{\^E}}1 {Î}{{\^I}}1 {Ô}{{\^O}}1 {Û}{{\^U}}1
  {Ã}{{\~A}}1 {ã}{{\~a}}1 {Õ}{{\~O}}1 {õ}{{\~o}}1
  {œ}{{\oe}}1 {Œ}{{\OE}}1 {æ}{{\ae}}1 {Æ}{{\AE}}1 {ß}{{\ss}}1
  {ű}{{\H{u}}}1 {Ű}{{\H{U}}}1 {ő}{{\H{o}}}1 {Ő}{{\H{O}}}1
  {ç}{{\c c}}1 {Ç}{{\c C}}1 {ø}{{\o}}1 {å}{{\r a}}1 {Å}{{\r A}}1
}

%% file: contents4/00_Abstract.tex
\begin{abstract}
Decentralized applications (dApps) in Decentralized Finance (DeFi) face a fundamental tension between regulatory compliance requirements like Know Your Customer (KYC) and maintaining decentralization and privacy. 
Existing \textit{permissioned DeFi} solutions often fail to adequately protect private attributes of dApp users and introduce implicit trust assumptions, undermining the blockchain's decentralization. 
Addressing these limitations, this paper presents a novel synthesis of Self-Sovereign Identity (SSI), Zero-Knowledge Proofs (ZKPs), and Attribute-Based Access Control to enable privacy-preserving on-chain permissioning based on decentralized policy decisions. 
We provide a comprehensive framework for permissioned dApps that aligns decentralized trust, privacy, and transparency, harmonizing blockchain principles with regulatory compliance.
Our framework supports multiple proof types (equality, range, membership, and time-dependent) with efficient proof generation through a commit-and-prove scheme that moves credential authenticity verification outside the ZKP circuit.
Experimental evaluation of our KYC-compliant DeFi implementation shows considerable performance improvement for different proof types compared to baseline approaches. 
We advance the state-of-the-art through a holistic approach, flexible proof mechanisms addressing diverse real-world requirements, and optimized proof generation enabling practical deployment.
\end{abstract}

%% file: contents4/01_Introduction.tex
\section{Introduction}
\label{sec:introduction}
Decentralized applications (dApps) running on public blockchains enable transparent, tamper-proof services without relying on centralized intermediaries.
This increases user control, reduces single points of failure, and facilitates trusted interactions.
Despite running on permissionless infrastructure, dApps must restrict access to users meeting certain conditions, such as holding minimum token balances, being whitelisted, or being a real human. 
This is driven by application-specific needs or external legal frameworks, such as Anti-Money Laundering (AML) and Know Your Customer (KYC) regulations in Decentralized Finance (DeFi).
On-chain KYC is typically only enforced when off-chain funds are transferred on-chain, or vice versa, leaving on-chain activities unchecked.
In this context, on-chain permissioning mitigates illicit activities and helps to fulfill regulatory obligations throughout the system.
However, the decentralized and permissionless nature of blockchains makes it challenging to apply concepts and analogies for permissioning as for centralized counterparts.
These challenges must be addressed in an integrated way.

A challenge arises from the need to implement transparent policy decisions without relying on centralized platforms.
On-chain permissioning, as proposed in~\cite{BloBaAcCoSe, BloAudAccConSys, SSIBloAccConSupAttPriZK} addresses this challenge by implementing permission verification directly through smart contracts.
This ensures that policy decisions are transparently recorded and verifiable.
However, this decentralized approach requires identity management where users maintain sovereignty over their credentials. 
Self-Sovereign Identity (SSI) facilitates this by giving users control over their credentials through personal identity wallets instead of relying on centralized platforms.
Realizing SSI-based permissioning on public blockchains raises privacy concerns, as sensitive user attributes must be protected. 
Zero-knowledge proofs (ZKPs) address this by allowing users to prove access rights without revealing private information.

The combination of SSI and ZKPs for on-chain permissioning can resolve the tension between privacy and the need to make policy decision and enforcement decentralized and transparent.
However, orchestrating interactions between smart contracts, SSI systems, and ZKP protocols can introduce vulnerabilities or increased operational costs if not handled properly.
While previous work explores privacy-preserving KYC/AML systems and ZKP-based permissioning, existing solutions fall short taking a holistic perspective and focus on subsets of the problem, such as decentralized policy storage with centralized~\cite{SSIBAC,philipp2024daxiot, DecPriUsiBlocProPerDat,ma2021attribute,bera2020designing} or intransparent~\cite{steichen2018blockchain} enforcement, missing decentralization of the policy decision~\cite{PriPreKYCIdeAccAllBlo, biryukov2018privacy} or policy information~\cite{xiong_regkyc_2025, baldimtsi2024zklogin}, missing decentralized or self-sovereign identity management~\cite{oh2025zkaml}, or privacy-preserving policy decisions without addressing practical challenges, such as flexible proof types, performance, and storage requirements~\cite{SSIBloAccConSupAttPriZK, pauwels_zkkyc_2022, kalbantner_zkp_2024, rathee2022zebra}.
To address these gaps, this paper makes three main contributions:
\begin{itemize}
    \item 
    First, we present a general model for privacy-preserving, on-chain permissioning that combines SSI and ZKPs with predicate-based policy decisions, specifically designed for public blockchain environments. 
    This model provides a theoretical foundation for understanding how these technologies can work together end-to-end for decentralized, private, and transparent permissioning.
    \item 
    Second, we demonstrate how this model can be instantiated for dApps running on public blockchains, by applying our generalized solution to a use case in the DeFi ecosystem, that enforces regulatory compliance to KYC/AML requirements in a liquidity pool.
    \item 
    Third, we provide guidelines for the practical realization of our model. 
    We propose a commit-and-prove scheme that aims to improve performance and lower storage requirements by checking authenticity of credentials outside of the ZKP circuit.
    Based on a prototypical implementation of our model, we conduct an experimental evaluation. 
\end{itemize}
This paper is organized as follows:
In Section \ref{sec:overview}, we describe the use case of permissioned decentralized finance to motivate our problem, and derive our design goals and threat model.
In Section \ref{sec:system_design} we gradually extend our existing system through an SSI-based and an ZKP-based approach to fulfill our design goals.
Based on that in Section \ref{sec:practical_realization} we explain the practical realization of our system design, for which an implementation and evaluation is provided in Section \ref{sec:evaluation}.

%% file: contents4/02_Preliminaries.tex
\section{Preliminaries}
\label{sec:preliminaries}
In this section, we first cover relevant background on SSI-based privacy preserving on-chain permissioning.
Next, we describe the current state of the art, related to our work.

\subsection{Background}

\subsubsection{KYC/AML for permissioned DeFi}
\label{subsubsec:kycaml_in_permissioned_defi}
Decentralized Finance (DeFi) describes a blockchain-enabled financial ecosystem that claims to operate without a central authority in an open, permissionless, and transparent manner~\cite{DeFiReview}.
It encompasses financial activities including lending, borrowing, and decentralized exchanges~\cite{DeFinCriAttDeFi}.
DeFi users retain full control of their assets and only need a wallet to interact with applications.
In traditional finance services, enforcement of Know Your Customer (KYC)/Anti-Money Laundering (AML) regulations aims to mitigate such behavior.
KYC is a process to identify and verify customers during the business relationship with a financial institution. 
AML is a set of laws, regulations, and procedures intended to prevent money laundering~\cite{SurKYCAMLForCryTx}.
Typically, KYC/AML requirements in DeFi are only enforced when off-chain funds are transferred on-chain or vice versa~\cite{xiong_regkyc_2025}. 
This gave rise to permissioned DeFi, where KYC providers maintain whitelists used on-chain to validate user permissions, as applied by protocols like Aave~\cite{AaveArc}.
This approach is described in more detail through our use case presented in Subsection~\ref{subsec:usecase}.

\subsubsection{Attribute-Based Access Control}
\label{subsubsec:abac-xacml}
The Attribute-Based Access Control (ABAC) model makes policy decisions based on attributes associated with subjects, objects, and environmental context~\cite{NISTABAC}.
We consider a subject requesting access to an object controlled by an owner through attribute-based policies, containing conditions.
Conditions are tuples with an attribute key, operator, and value.
The ABAC model is conceptualized through the eXtensible Access Control Markup Language (XACML)~\cite{XACMLSpec}.
Our system comprises the following XACML components:
The Policy Retrieval Point (PRP) stores policies.
The Policy Decision Point (PDP) retrieves policies from the PRP, evaluates them, and returns decisions.
The Policy Enforcement Point (PEP) intercepts access requests and enforces PDP decisions.
The Policy Administration Point (PAP) creates and stores policies via the PRP.
The Policy Information Point (PIP) retrieves and updates attributes~\cite{XACMLRiskAwareAC}.

\subsubsection{Self-sovereign identity}
\label{subsubsec:ssi}
Self-sovereign identity (SSI) empowers users by granting them greater autonomy in managing their digital identities, while providing trust and decentralization~\cite{ABACAuthInfraECom}.
The SSI model works around the triangle-of-trust~\cite{ToPoQuZeKnVeCrSySeSoId}, defining three roles:
The holder maintains Verifiable Credentials (VCs) and creates Verifiable Presentations (VPs) for requesting a service or a resource from a verifier.
The issuer creates VCs by asserting claims on subjects and forwards them to holders.
The verifier receives and processes VPs~\cite{ToPoQuZeKnVeCrSySeSoId}.
A DID~\cite{DIDSpec} represents a unique digital identity in a decentralized context, associating a DID subject with a DID document containing public cryptographic material and verification methods~\cite{DIDSpec}.
A VC~\cite{VCSpec} contains a set of attributes assigned to a DID.
The VC is digitally signed by the issuer to attest the validity of its claims.
The VC schema~\cite{VCSchemaSpec} defines the specific fields, data types, and validation rules for a VC in a structured format.
To reveal a subset of their VCs to a verifier the subject creates a VP\@.
The VP creation process may be initialized by a verifier through a Verifiable Presentation Request (VPR)~\cite{VPRSpec}.
A VDR represents the underlying infrastructure supporting the issuance of DIDs and VCs by storing and resolving verification material, and other relevant data such as the DID document and VC schema.
Trusted databases, decentralized networks such as IPFS and distributed ledgers such as blockchains represent examples of a VDR~\cite{VCSpec}.

\subsubsection{Zero-knowledge succinct non-interactive arguments of knowledge}
\label{subsubsec:zksnarks}
Zero-knowledge succinct non-interactive arguments of knowledge (zk-SNARKs) are a class of non-interactive zero-knowledge protocols that distinguish through a succinct proof size and fast verification time~\cite{WeaZerKnoBeyBlaBoxPan}.
Among others, zk-SNARKs provide means for realizing Verifiable Off Chain Computation (VOC) by offloading computations off-chain without compromising their integrity~\cite{VOC}.
We use zk-SNARKs by describing three operations \( ZKSetup \), \( ZKProve \) and \( ZKVerify \):
\begin{itemize}
    \item \( ZKSetup(ecs, srs) \): 
    Takes as input an executable constraint systems \( ecs \) and a structured reference string \( srs \).
    It returns a proving key \( pk^{ZK} \) and a verification key \( vk^{ZK} \)
    \item \( ZKProve(ecs, x, x^{\prime}, w) \): 
    Takes as input the executable constraint system \( ecs \), a public input \( x \), a private input \( x^{\prime} \), a witness \( w \), and returns a proof \( \pi \)
    \item \( ZKVerify(x, vk^{ZK}, \pi) \): 
    Takes as input a public statement \( x \), the verification key \( vk^{ZK} \) and a proof \( \pi \).
    It returns \( 1 \), if \( \pi \) combined with \( vk^{ZK} \) successfully verifies against \( x \), otherwise \( 0 \)
\end{itemize}

\subsection{Related work}
\label{subsec:related_work}
We identified related works in three areas: on-chain KYC/AML, on-chain permissioning, and ZKPs and credentials.
Many works have suggested approaches to compromise the tension between regulatory concerns and privacy on the blockchain. %
In \cite{mansoor_review_2023, hannan_systematic_2023, HowBloAutKYC, kumar2020blockchain}, approaches for \textit{on-chain KYC/AML} are proposed or evaluated that focus on the applicability in traditional finance (TradFi) to make it less time consuming and more cost efficient.
In \cite{xiong_regkyc_2025, PriPreKYCIdeAccAllBlo, pauwels_zkkyc_2022, biryukov2018privacy, oh2025zkaml, kalbantner_zkp_2024}, approaches for privacy-preserving on-chain KYC/AML using ZKPs are proposed.
These works address specific aspects rather than taking a holistic perspective aligned with our design goals:
\cite{PriPreKYCIdeAccAllBlo, biryukov2018privacy} focus on centralized policy decisions, \cite{xiong_regkyc_2025} does not address decentralized policy information, \cite{oh2025zkaml} does not cover self-sovereign identity management, and \cite{pauwels_zkkyc_2022, kalbantner_zkp_2024} primarily address privacy-preserving decisions without considering practical challenges like flexible proof types, performance, and storage requirements.

Works on \textit{on-chain permissioning} either execute policy decisions on-chain for off-chain enforcement~\cite{SSIBAC, philipp2024daxiot, steichen2018blockchain, DecPriUsiBlocProPerDat, ma2021attribute, bera2020designing} or use zk-SNARKs for on-chain enforcement~\cite{SSIBloAccConSupAttPriZK, rathee2022zebra, baldimtsi2024zklogin}.
The former often rely on centralized enforcement\cite{SSIBAC, philipp2024daxiot, DecPriUsiBlocProPerDat, ma2021attribute, bera2020designing} or lack transparency~\cite{steichen2018blockchain}, while the latter tend to overlook practical implementation considerations.
Our work addresses practical considerations, by introducing a commit-and-prove scheme to reduce zk-SNARKs runtimes and storage requirements, characterizing policy decisions through multiple proof types such as membership and time-dependent proofs, and incorporating a \( VPR^{ZK} \) that enables a machine-readable \( VP^{ZK} \).
Beyond academic research, several projects support on-chain permissioning, most notably OpenZeppelin~\cite{OpenZeppelin}.
Other non-academic projects such as iden3~\cite{Iden3}, Privado ID~\cite{PrivadoID}, and Semaphore~\cite{Semaphore} enable privacy-preserving on-chain permissioning using zkSNARKs.

Many works investigate the use of \textit{ZKPs and credentials}.
This applies to both verifiable credentials in the SSI context and anonymous credentials.
They improve upon the expressiveness of proofs~\cite{NonDisCreOncBlocDApps, rosenberg_zk-creds_2023, muth_AnonCreds_2023, camenisch2015composable, connolly2022improved, heiss_DePINs_2024}, the speed of proving and verification, which can be prohibitive for practical on-chain verification~\cite{rathee2022zebra, hanzlik2021little}, or different levels of accountability and revocation~\cite{sonnino_coconut_2019, belenkiy2008p, camenisch2001efficient, rathee2022zebra}.
Compared to our work, there is less focus on practical on-chain integration and regulatory requirements~\cite{NonDisCreOncBlocDApps, rosenberg_zk-creds_2023, muth_AnonCreds_2023, connolly2022improved, heiss_DePINs_2024}, a centralized policy enforcement\cite{rathee2022zebra, hanzlik2021little}, or less focus on a decentralized policy management~\cite{sonnino_coconut_2019, belenkiy2008p, camenisch2001efficient}.

%% file: contents4/03_Model.tex
\section{Model}
\label{sec:overview}
In this section, we first motivate the need for privacy-preserving on-chain policy decisions for KYC/AML in permissioned DeFi by the example of liquidity pools. 
Based on that, we derive our threat model and design goals.

\subsection{Use case}
\label{subsec:usecase}
\begin{figure}[htbp]
    \centering
    \includegraphics[width=0.8\columnwidth]{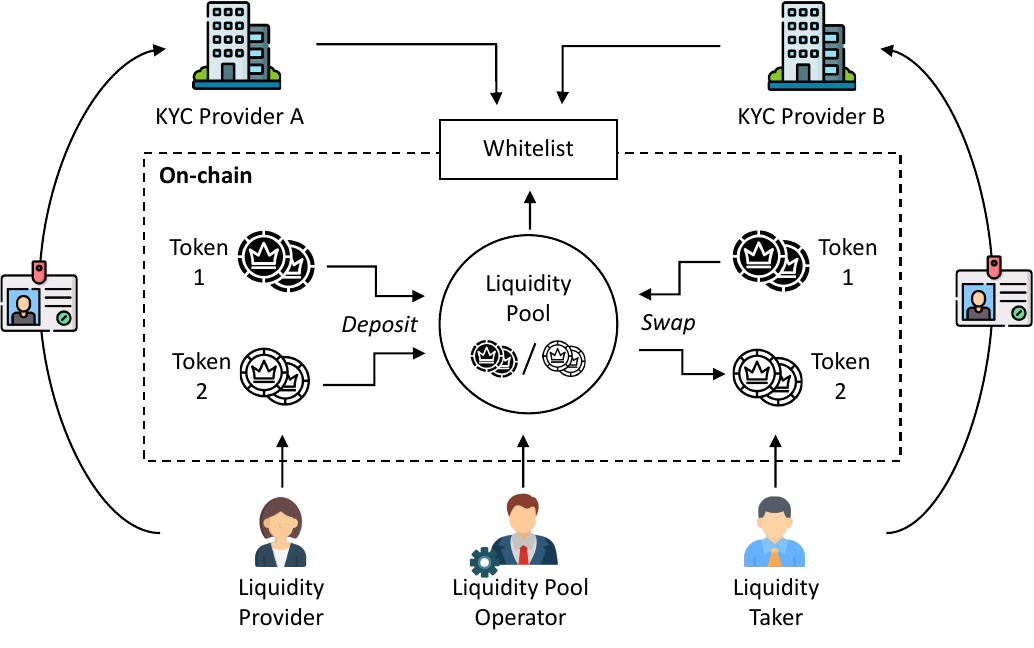}
    \caption{KYC for a permissioned liquidity pool}
    \label{fig:kyc}
\end{figure}

For the use case of this paper, we consider permissioned liquidity pools that are accessible to any user that is compliant with KYC/AML regulations.
Figure \ref{fig:kyc} shows this approach.
It involves the entities KYC provider, liquidity provider, liquidity taker and liquidity pool operator.
KYC providers are responsible for verifying the identities of users according to KYC/AML standards. 
Liquidity providers contribute assets such as Token 1 and Token 2 to the liquidity pool in exchange for interest.
Liquidity takers, interact with the pool by swapping assets. 
All participants in the pool must have gone through a KYC process with a KYC provider approved by the liquidity pool operator. 
This process typically includes collecting identity documents, screening against sanction lists, and assessing the risk level of a user.
Once verified, their blockchain address is added to a whitelist.
It is enforced by the liquidity pool smart contract before executing a deposit or swap.

This introduces the following challenges:
The KYC provider determines if and how access is granted (i.e., if attributes are valid).
This leads to single points of failure and centralized control over policy decisions.
The system is unable to protect personal information because each KYC provider needs to view all of a user's attributes (e.g., all attributes contained in an identity document), even though only a subset is required for the policy decision.
This sensitive information could even be leaked if the KYC provider suffers a data breach.
Last, the policy decision is not transparent.
Attributes are verified off-chain, and only the verification result is stored on-chain.
Users of the liquidity pool are unable to observe if policy decisions were made according to the defined policies.


\subsection{Threat model}
\label{subsec:threat_model}
In order to describe the threat model, we first define the roles of a generalized model derived from the permissioned DeFi use case.
These roles include the subject, owner, and issuer.
The subject holds multiple attributes in a VC, and requests to execute a function (\( f_{id} \)) in a smart contract with a unique identifier (\( id \)) for that function.
The owner controls the smart contract function.
It can be represented by multiple identities, for example, a decentralized autonomous organization (DAO), and is typically the entity that developed the protocol and deployed the smart contracts.
The issuer operates according to the SSI model.
It acts as a trusted third party, certifying that the subject holds certain attributes by issuing VCs.

The subject may act maliciously.
It may attempt to provide tampered attributes, steal authentication information, or forge authentication.
Owners are not trustworthy.
They may set up misconfigured policies or unduly deny or allow access to the smart contract functions they control.
They may also be interested in violating the confidentiality of a subject's attributes.
According to the SSI-based trust model, the issuer is considered legitimate by both subjects and owners.
VCs attested by the issuer are considered true by all participants.
If an entity does not trust the issuer, it will not involve them in its operations. 
For example, the subject will not hold VCs from such an issuer, and the owner will not rely on them.

\subsection{Design goals}
\label{subsec:design_goals}
We aim to gradually design a system for on-chain permissioning for dApps that achieves the following design goals:
\begin{itemize}
    \item \textbf{Decentralized trust:}
    As described in Section~\ref{subsec:related_work} existing solutions often rely on centralized components for policy decisions or enforcement.
    Our system should support decentralized trust for all XACML components presented in Subsection~\ref{subsubsec:abac-xacml}.
    For a decentralized PIP, our goal is to design a system where identity attributes are self-sovereign, and subjects control their credentials while leveraging attestations from multiple issuers.
    For a decentralized PEP, PDP, and PRP, single points of failure and manipulation by the owner should be prevented.
    \item \textbf{Privacy preservation:}
    Following data minimization principles, our aspired system design uses selective disclosure and cryptographic mechanisms to support on-chain policy decisions while protecting personal information.
    \item \textbf{Transparency:}
    We aim to design a system, that records policy definitions and enforcement actions on-chain, using cryptography to balance auditability with privacy. 
    This enables compliance verification while maintaining confidentiality of sensitive information.
\end{itemize}

%% file: contents4/04_System_Design.tex
\section{System design}
\label{sec:system-design}
\label{sec:system_design}
Addressing the previously identified challenges, we first present a decentralized system for on-chain permissioning synthesizing SSI and ABAC. Then, we extend the design through zk-SNARKs to add privacy.

\subsection{SSI-based on-chain permissioning}
\label{sec:self-sovereign-identity}
A generalized system design for SSI-based on-chain permissioning assigns privileges based on attributes.
This approach is naturally reflected in the ABAC model, which is conceptualized through XACML.
The subject, owner, and issuer are defined as described in Subsection~\ref{subsec:threat_model}.
They are identified by their DID.
The DID, VC, VP, and VPR are described in Subsubsection~\ref{subsubsec:ssi}.
A VC contains claims about the subject, attested by the issuer.
The subject holds one or more VCs and creates the VP.
The VPR is created by the owner and describes how to authenticate the subject and which VCs should be included to grant access according to the policy.
A policy is defined as a mapping of \( id \) to a VPR, that defines the conditions to meet in order to execute the smart contract function.

\begin{figure}[htbp]
    \centering 
    \includegraphics[width=\columnwidth]{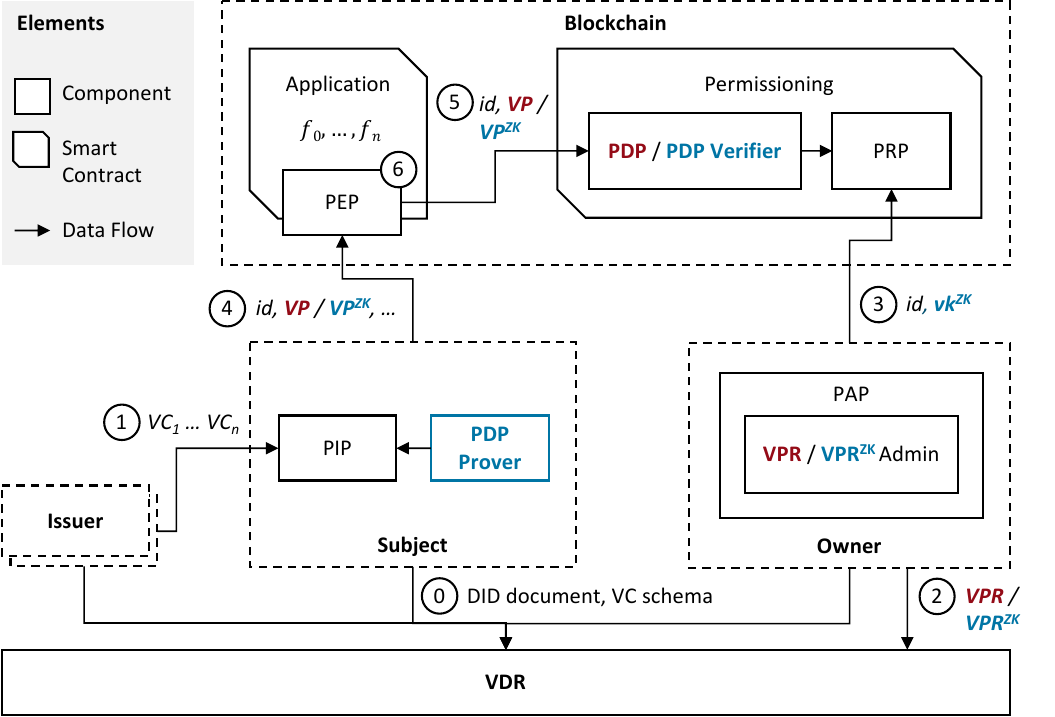}
    \caption{The flow for realizing SSI-based on-chain permissioning. Blue / red elements highlight a flow with / without privacy.}
    \label{fig:abacssi}
\end{figure}

The VDR component is introduced to support the storage of issuance details specific to DIDs and VCs.
The system comprises the XACML components PRP, PDP, PEP, PAP, and PIP.
The application smart contract holds the application logic
and enforce policies by interacting with the PDP.
The permissioning smart contract includes the PRP for storing policies and the PDP for evaluating policy decisions.
The PIP moves to the subject, which is responsible for the maintenance of its own attributes as a VC.
As a component of the owner's PAP, the VPR Admin is introduced.
It contains the required functionalities for managing a VPR.
Figure \ref{fig:abacssi} shows this flow.
Blue elements highlight the flow for SSI-based on-chain permissioning.
The interactions are as follows:
As a prerequisite, step \step{0} involves the issuer creating and storing a VC schema in the VDR. 
Additionally, the owner, subject, and issuer each create a DID by generating a key pair and uploading their respective DID documents to the VDR.
In \step{1}, the issuer creates a VC by referencing the VC schema and attesting to the subject's attributes. 
The VC is then sent to the subject.
In \step{2}, the owner's VPR admin creates a VPR using a set of conditions and the \( id \) as input, based on the VC schema, and stores it publicly in the VDR.
In \step{3}, the owner creates a policy using \( id \) and stores it in the PRP within the permissioning smart contract, referencing the VPR from step \step{2}.
In \step{4}, the subject creates a VP based on the VPR and sends it to a function \( f_{id} \) of the application smart contract.
In \step{5}, the application smart contract's PEP forwards the transaction to the on-chain PDP. 
The PDP retrieves the associated policy from the PRP and evaluates the policy decision by verifying the VP. 
This verification ensures that: (1) claims in the VP are attested by trusted issuers, (2) claims correspond to conditions C defined in the VPR, and (3) the subject successfully authenticates.
Last, the PEP executes the transaction by calling \( f_{id} \) when access is granted.

\subsection{Adding Privacy}
\label{subsec:abac-ssi-zkp}
Combining SSI with ZKPs allows a subject to prove access without disclosing the DID or VC to the owner~\cite{ZKProofOfIdentity}. 
This is achieved by leveraging zk-SNARKs.

The following changes are introduced and highlighted in red in Figure \ref{fig:abacssi}:
The VP becomes a zero-knowledge verifiable presentation \( VP^{ZK} \) containing the returned proof \( \pi \) of \( ZKProve \).
It can be presented without violating the privacy of a VC.
The VPR becomes a zero-knowledge verifiable presentation request \( VPR^{ZK} \) extending VPR with parameters for \( ZKProve \), including \( pk^{ZK} \), and \( ecs \).
To distinguish between the zk-SNARKs proving and verification, the PDP functionality is split into a PDP verifier and a PDP prover.
The \( VPR^{ZK} \) admin extends the VPR admin by capabilities for managing \( VPR^{ZK} \).
In Figure~\ref{fig:abacssi}, the interactions for \step{0}, \step{1}, and \step{2} stay unchanged.
Step \step{3} is extended to support the zk-SNARKs proving and verification.
Here, inputs are modified to take a \( VPR^{ZK} \) instead of a VPR.
This reflects the change of VPR to \( VPR^{ZK} \) which is now performing \( ZKSetup \) as well as creating and registering \( VPR^{ZK} \) in VDR.
The parameters \( id \) and \( VPR^{ZK} \) are written to PRP.
Step \step{4} is modified to support the generation of \( VP^{ZK} \).
Therefore, it entails performing \( ZKProve \) with the content of VC as private input \( x^{\prime} \).
A detailed explanation of the proof generation is given in Section~\ref{sec:practical_realization}.
Step \step{5} is modified to include the required operations for performing a zk-SNARKs verification.
It includes \( ZKVerify \) where the public parameters \( x \) are optional and provided by the owner, \( vk^{ZK} \) is retrieved from the PRP of the owner and \( \pi \) is included in \( VP^{ZK} \) sent by the subject in the access request.
Step \step{6} remains unchanged.

%% file: contents4/05_Practical_Realization.tex
\section{Practical realization}
\label{sec:practical_realization}
We now describe how the system design for privacy-preserving on-chain permissioning can be put into practice by highlighting the characteristics of its realization.

\subsection{Main idea}
We realize a privacy-preserving on-chain policy decision, by verifying a proof for the correct execution of the following three operations: 
(i) The \textbf{key binding} ensures that a subject is the holder of the VC, (ii) the \textbf{authenticity check} ensures that the correct issuer has attested the VC, and (iii) the \textbf{compliance check} ensures that the VC corresponds to the conditions specified by the policy.
Based on that, we define a set of claims as \( CL = \{ cl_1 \dots cl_n \} \) with \( cl \) representing a claim that contains a tuple of \( ( pk_{\sdv}, attr ) \) with the public key of a subject and an attribute \( attr \), containing an attribute key and attribute value as \( attr = (key, value) \).
The verifiable credential VC is defined as \( VC = ( pk_{\idv}, CL, \sigma_{CL} ) \) with \( \sigma_{CL} = DSSign(sk^{DI}_{\idv}, CL) \).
We use digital signatures to authenticate entities by describing two operations \( DSSign \) and \( DSVerify \):
\begin{itemize}
    \item \( DSSign(sk_\hdv, m) \):
    Takes as input the private key of a holder \( sk^{DI}_\hdv \) and a message \( m \).
    It returns the signature \( \sigma_m \) over \( m \).
    \item \( DSVerify(pk_\hdv, \sigma_m, m) \):
    Takes as input the public key of a holder \( pk_\hdv \), a signature \( \sigma_m \), a message \( m \) and outputs \( 1 \) if \( m \) was signed using \( sk_\hdv \) corresponding to \( pk_\hdv \) and \( 0 \) otherwise.
\end{itemize}
We define the compliance check by the operation \( CompCheck \):
\begin{itemize}
    \item \( CompCheck(attr, c, aux) \):
    Takes as input a condition \( c \), an attribute \( attr \), and auxiliary data \( aux \).
    It returns \( true \) if \( attr \) was successfully evaluated against \( c \) using \( aux \) and \( false \) otherwise.
\end{itemize}
The operation \( CompCheck \) embeds the logic for assuring compliance to different conditions.
The conditions are defined by the proof type as described in Subsection \ref{subsec:proof_types}.

A naive approach performs the key binding, authenticity check, and compliance check as part of the circuit. 
When verifying a signature, such as for the key binding and authenticity checks, the verifier first computes the hash and then verifies the signature's validity over the resulting hash digest. 
We take a commit-and-prove scheme similar to that proposed in \cite{EffVerImaRed, VerITAS}, which allows the authenticity check to be performed outside the circuit.
The key binding and compliance checks, however, still remain in the circuit. 
This approach requires only a single signature verification for key binding inside the circuit instead of multiple verifications per claim, as opposed to the authenticity check.
The authenticity check is performed on-chain.
The key binding is still executed within the circuit to keep the subject's public key hidden on-chain.

\subsection{Proof generation}
A proof for key binding and compliance is generated by the subject and represents the zero-knowledge verifiable presentation \( VP^{ZK} \).
A zero-knowledge verifiable presentation request \( VPR^{ZK} \) describes the criteria for creating the \( VP^{ZK} \).
It defines the set of conditions \( C \), the executable constraint system \( ecs \), as well as the zk-specific proving-\/and verification keys \( pk^{ZK} \) and \( vk^{ZK}\).
The problem logic for evaluating a policy in a provable representation is encoded in \( ecs \) and shown in Figure~\ref{fig:proving_program}.
\begin{algorithm}
    \label{fig:proving_program}
    \caption{Circuit specific policy decision logic} 
    \hspace*{\algorithmicindent} \textbf{Public Input:} \( \eta, H, aux \) \\
    \hspace*{\algorithmicindent} \textbf{Private Input} \( VC, pk_{\sdv}, \sigma_{\eta} \)
    \begin{algorithmic}[1]
        \State {\( ( pk_\idv, CL, \sigma_{CL} ) := VC \)}
        \State {\( DSVerify(pk_\sdv, \sigma_{\eta}, \eta) \)}
	\For {\(i \in \{1, \dots, length(CL) \}\)}
            \State {\( ({pk_\sdv}^{\prime}, attr) := CL_i \)}
            \State {\( assert(pk_\sdv = {pk_\sdv}^{\prime} ) \)}
            \State {\( assert(H_i = hash(CL_i) ) \)}
            \State {\( CompCheck(attr, C_i, aux) \)}
	\EndFor
    \end{algorithmic}
    \label{alg:policy_decision}
\end{algorithm}
It takes as public inputs a nonce \( \eta \), a set of hashes \( H \), the set of conditions \( C \), auxiliary data \( aux \) and as private inputs the VC, the subject public key \(  pk_{\sdv} \) and the signature of the subject over the nonce \( \sigma_n \).
In Line 2 the key binding is checked by executing \( DSVerify \) on \( \sigma_n \) for \( pk_{\sdv} \).
In Line 5, for every claim \( CL_i \) in VC, correct ownership of the claim is ensured by validating equality of the subject public key \( pk_{\sdv} \) passed as the public parameter with the subject public key \( pk^{\prime}_{\sdv} \) included in the claim.
Line 6 ensures equality between the computed hash of a claim with the hash \( H_i \) passed as public input.
It assumes authenticity was checked on-chain by performing \( DSVerify(pk_{\idv}, \sigma_{H_i}, H_i) \).
In Line 7 the compliance check is performed by executing \( CompCheck \) for the attribute \( attr \) of \( CL_i \) in VC with \( C_i \) and the optional parameter \( aux \).
\( CompCheck \) executes the condition depending on the proof type as described in the following Subsection~\ref{subsec:proof_types}.

\subsection{Proof types}
\label{subsec:proof_types}
The proposed realization allows for defining different types of conditions, which are embedded in the logic of \( CompCheck \) and define the type of proof.
They can be used either on their own or in combination with each other.
A detailed explanation of this approach is given in \cite{NonDisCreOncBlocDApps}.
The proof types are defined as follows:
\begin{itemize}
    \item Range proofs: 
    Define that a numeric attribute must be in a certain interval.
    In Algorithm~\ref{alg:policy_decision}, this involves \( c \) to consist of a tuple \( (key, operator, value) \) with \( operator \in \{ >, <, \geq, \leq \} \).
    \item Equality proofs: 
    Define that an attribute must be equal to a value.
    In Algorithm~\ref{alg:policy_decision}, this involves \( c \) to consist of a tuple \( (key, operator, value) \) with \( operator \in \{ =, \neq \} \).
    \item Membership proofs: 
    Define if an attribute is in a predefined set using a Merkle tree structure to allow for non-disclosing membership proofs.
    In Algorithm~\ref{alg:policy_decision}, this involves \( c \) to consist of a tuple \( (key, operator, value) \) with \( operation \in \{ in, notin \} \). 
    Furthermore, as auxiliary input \( aux \), the set as well as the root of the Merkle tree must be provided.
    \item Time-dependent proofs: 
    Define if a date- or time-based attribute is in a range that has boundaries relative to the current date or a timestamp.
    They are specified for a range or equality proof using \( value \) with \( c = (key, operator, value) \) as the current timestamp.
\end{itemize}

%% file: contents4/06_Evaluation.tex
\section{Evaluation}
\label{sec:evaluation}
We demonstrate technical feasibility of the proposed system through a prototypical implementation and provide first insights into its practicality by conducting initial experiments.

\subsection{Use case revisited}
Based on Section~\ref{subsec:usecase}, we instantiate our use case as follows:
The liquidity provider and liquidity taker represent the subject seeking access to smart contract functions for depositing tokens \( id_{deposit} \) and swapping tokens \( id_{swap} \).
The liquidity pool operator represents the owner, deploying and operating the pool while defining KYC/AML compliance conditions.
For simplicity, this role is controlled by a single identity.
It can also be represented by multiple identities or a DAO.
A federal technology company acts as the issuer, attesting identity documents as VCs containing a unique identifier, date of birth, and address.
The KYC/AML regulations align with the proof types from Subsection~\ref{subsec:proof_types}:
First, users must reside in a specific country, ensured by an equality proof matching the VC address to a specified value.
Second, users must be at least 18 years old, verified using a time-dependent proof with the current block time as a public parameter.
Third, users must not appear on sanctioned lists, verified using membership proof based on the VC's unique identifier.
These conditions are encoded in \( VPR^{ZK} \) logic uploaded to a VDR (e.g., IPFS) and anchored on-chain, referencing \( id_{deposit} \) and \( id_{swap} \).
For depositing, the liquidity provider uses its VC to build \( VP^{ZK} \) in the PDP prover as described by \( VPR^{ZK} \) for \( id_{deposit} \), attaching \( VP^{ZK} \) for verification during smart contract execution.
For swapping, the liquidity taker follows similar steps using its VC to build and attach \( VP^{ZK} \) for \( id_{swap} \).

\subsection{Implementation}
The proposed system has been implemented by providing multiple script files written in Python.
The source code is available on Github\footnote{https://github.com/fapiper/zk-ssi-onchain-permissioning}
The smart contracts are implemented in Solidity\footnote{https://soliditylang.org}.
Step \step{2} is implemented in Python using zokrates\_pycrypto.
The signature employs the Edwards-curve Digital Signature Algorithm (EdDSA) scheme and the SHA256 hash function.
The \( ecs \) for \( ZKSetup \) in \step{3} is implemented using the ZoKrates domain-specific language.
It can be compiled using the ZoKrates command line interface which is used for creation of \( pk^{ZK} \) and \( vk^{ZK} \).
Step \step{4} is realized using two ZoKrates-specific phases.
First, during the witness phase the input-specific variable assignment of the \( ecs \) is computed.
Next, during the proving phase the \( VP^{ZK} \) is built.
For realizing \( ZKVerify \) in \step{5}, the function \( verifyTx() \) of a smart contract is used.
It is generated by the ZoKrates toolbox and already includes \( vk^{ZK} \).

\subsection{Experiments}
\label{subsec:experiments}
To gain first insights in the practicality of the system and to explore the potential of performing the authenticity check outside the ZKP circuit, we execute a first set of experiments on our prototype.

\subsubsection{Experiment design}
Using zk-SNARKs usually comes with significant drawbacks in terms of performance.
In our presented solution, these are associated with the execution of \step{3}, \step{4} and \step{5}, that require the execution of the zk-SNARKs operations \( ZKSetup \), \( ZKProve \) and \( ZKVerify\).
Executing \( ZKVerify\) in \step{5} on-chain is associated with transaction costs. 
They are considered as the main bottleneck of our system and are expected to have the greatest impact on the practicability of our solution.
To achieve these objectives, we analyze the behavior of the system for different workloads, considering 1, 2, 4, 8, and 16 conditions. 
We compare the results of the commit-and-prove approach, when performing the authenticity check outside the ZKP circuit (\textit{cp-*}), with a baseline approach where the authenticity check is performed inside the circuit.
Referring to the proof types presented in Section \ref{subsec:abac-ssi-zkp}, we provide measurements on the equality type (\textit{equal}), range type (\textit{range}) and membership type with a Merkle tree depth of 3 (\textit{member}).
Time-based proofs involve range proofs and are not taken into account.
The experiments are performed using an Intel Core i7-11800H@2.3 GHz machine with 16 cores and 16GB RAM, measuring artifact sizes of \( Compiled \) and \( pk^{ZK} \) in megabytes (\unit{\mega\byte}), \( vk^{ZK} \) in kilobytes (\unit{\kilo\byte}), runtimes of witness generation, setup, proof generation in seconds (\unit\second), and the transaction costs of verifying the proof in gas (\( gas \))~\cite{ETHYellow}.
 
\subsubsection{Experiment results}
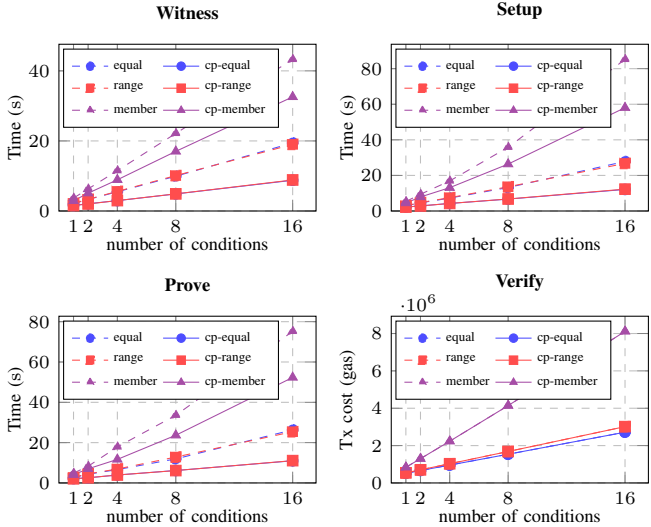
\begin{figure}[htbp]
    \centering
    \begin{tikzpicture}
        \begin{groupplot}[
          group style={group size=2 by 2, vertical sep=1.4cm},
          x label style={at={(axis description cs:0.5,-0.1)},anchor=north},
          y label style={at={(axis description cs:-0.1,0.5)},anchor=south}, 
          xlabel={number of conditions},
          ylabel={Time (\unit\second)},
          height=3.8cm, width=5cm,
          grid style=dashed,
          ymin=0,
          xmin=0,
          xtick={1,2,4,8,16},
          ymajorgrids=true,
          xmajorgrids=true,
          /tikz/font =\scriptsize,
          legend cell align=left,
          legend image post style={scale=0.8},
          legend columns=2,
          legend style={font=\tiny, at={(0.436,0.974)}, anchor=north}
        ]
        \nextgroupplot[title=\textbf{Witness}]
        \addplot[dashed,blue!70,mark=*] 
        coordinates {
            (1, 2.12)
            (2, 3.29)
            (4, 5.49)
            (8, 9.93)
            (16, 19.44)
        };
        \addlegendentry{equal};
        \addplot[blue!70,mark=*] 
        coordinates {
            (1, 1.59)
            (2, 2.07)
            (4, 2.96)
            (8, 4.87)
            (16, 8.77)
        };
        \addlegendentry{cp-equal};
        \addplot[dashed,red!70,mark=square*]
            coordinates {
            (1, 2.18)
            (2, 3.34)
            (4, 5.60)
            (8, 10.11)
            (16, 19.02)
        };
        \addlegendentry{range};
        \addplot[red!70,mark=square*]
            coordinates {
            (1, 1.55)
            (2, 2.01)
            (4, 2.97)
            (8, 4.89)
            (16, 8.86)
        };
        \addlegendentry{cp-range};
        \addplot[dashed,violet!70,mark=triangle*]
        coordinates {
            (1, 3.69)
            (2, 6.30)
            (4, 11.52)
            (8, 22.15)
            (16, 43.27)
        };
        \addlegendentry{member};
        \addplot[violet!70,mark=triangle*]
        coordinates {
            (1, 2.90)
            (2, 5.09)
            (4, 8.91)
            (8, 17.02)
            (16, 32.58)
        };
        \addlegendentry{cp-member};
        \nextgroupplot[title=\textbf{Setup}]
        \addplot[dashed,blue!70,mark=*] 
        coordinates {
            (1, 2.94)
            (2, 4.57)
            (4, 7.27)
            (8, 13.20)
            (16, 27.84)
        };
        \addlegendentry{equal};
        \addplot[blue!70,mark=*] 
        coordinates {
            (1, 2.26)
            (2, 2.87)
            (4, 4.24)
            (8, 6.69)
            (16, 12.12)
        };
        \addlegendentry{cp-equal};
        \addplot[dashed,red!70,mark=square*]
            coordinates {
            (1, 2.95)
            (2, 4.74)
            (4, 7.49)
            (8, 13.62)
            (16, 26.74)
        };
        \addlegendentry{range};
        \addplot[red!70,mark=square*]
            coordinates {
            (1, 2.20)
            (2, 2.82)
            (4, 4.25)
            (8, 6.75)
            (16, 12.29)
        };
        \addlegendentry{cp-range};
        \addplot[dashed,violet!70,mark=triangle*]
        coordinates {
            (1, 5.33)
            (2, 9.41)
            (4, 16.89)
            (8, 35.81)
            (16, 85.20)
        };
        \addlegendentry{member};
        \addplot[violet!70,mark=triangle*]
        coordinates {
            (1, 4.48)
            (2, 7.90)
            (4, 13.11)
            (8, 26.37)
            (16, 58.07)
        };
        \addlegendentry{cp-member};
        \nextgroupplot[title=\textbf{Prove}]
        \addplot[dashed,blue!70,mark=*] 
        coordinates {
            (1, 2.74)
            (2, 4.33)
            (4, 6.63)
            (8, 11.93)
            (16, 26.35)
        };
        \addlegendentry{equal};
        \addplot[blue!70,mark=*]
        coordinates {
            (1, 2.14)
            (2, 2.57)
            (4, 3.94)
            (8, 6.17)
            (16, 10.99)
        };
        \addlegendentry{cp-equal};
        \addplot[dashed,red!70,mark=square*]
        coordinates {
            (1, 2.74)
            (2, 4.51)
            (4, 6.92)
            (8, 12.85)
            (16, 25.42)
        };
        \addlegendentry{range};
        \addplot[red!70,mark=square*]
        coordinates {
            (1, 2.06)
            (2, 2.56)
            (4, 3.94)
            (8, 6.14)
            (16, 11.07)
        };
        \addlegendentry{cp-range};
        \addplot[dashed,violet!70,mark=triangle*]
        coordinates {
            (1, 4.87)
            (2, 8.29)
            (4, 17.86)
            (8, 33.57)
            (16, 75.19)
        };
        \addlegendentry{member};
        \addplot[violet!70,mark=triangle*]
        coordinates {
            (1, 4.07)
            (2, 7.01)
            (4, 11.68)
            (8, 23.66)
            (16, 52.34)
        };
        \addlegendentry{cp-member};
        \nextgroupplot[title=\textbf{Verify}, xlabel={number of conditions},
          ylabel={Tx cost (gas)}]
        \addplot[dashed,blue!70,mark=*] 
        coordinates {
            (1, 522731)
            (2, 666779)
            (4, 955882)
            (8, 1537941)
            (16, 2717635)
        };
        \addlegendentry{equal};
        \addplot[blue!70,mark=*] 
        coordinates {
            (1, 522707)
            (2, 666779)
            (4, 955858)
            (8, 1537953)
            (16, 2717635)
        };
        \addlegendentry{cp-equal};
        \addplot[dashed,red!70,mark=square*]
            coordinates {
            (1, 540673)
            (2, 702736)
            (4, 1028141)
            (8, 1683897)
            (16, 3015043)
        };
        \addlegendentry{range};
        \addplot[red!70,mark=square*]
            coordinates {
            (1, 540637)
            (2, 702736)
            (4, 1028153) 
            (8, 1683885)
            (16, 3015031)
        };
        \addlegendentry{cp-range};
        \addplot[dashed,violet!70,mark=triangle*]
        coordinates {
            (1, 837790)
            (2, 1299877)
            (4, 2233915)
            (8, 4141530)
            (16, 8114710)
        };
        \addlegendentry{member};
        \addplot[violet!70,mark=triangle*]
        coordinates {
            (1, 837790)
            (2, 1299877)
            (4, 2233915)
            (8, 4141518)
            (16, 8114734)
        };
        \addlegendentry{cp-member};
        \end{groupplot}
    \end{tikzpicture}
     \caption{Performance of the witness, setup, and prove phases, as well as transaction (tx) costs of the verification for equality, range, and membership proofs with respect to the number of conditions.}
    \label{fig:performance}
\end{figure}

\begin{figure*}[htbp]
    \centering
    \begin{tikzpicture}
        \begin{groupplot}[
          ybar,
          /pgf/bar width=3pt,
          ybar=1pt,
          xtick=data,
          ymin=0,
          symbolic x coords={1,2,4,8,16},
          enlarge x limits=0.14,
          group style={group size=3 by 1, horizontal sep=1.2cm},
          x label style={at={(axis description cs:0.5,-0.18)},anchor=north},
          y label style={at={(axis description cs:-0.1,0.5)},anchor=south},
          xlabel={number of conditions},
          ylabel={Size (\unit{\kilo\byte})},
          height=3.5cm, width=6.4cm,
          grid style=dashed, 
          ymajorgrids=true,
          xmajorgrids=true,
          /tikz/font =\scriptsize,
          legend cell align=left,
          legend columns=2,
          legend style={font=\tiny,at={(0.28,0.97)}, anchor=north}
        ]
        \nextgroupplot[title=\textbf{Compiled}, ylabel={Size (\unit{\mega\byte})}]
        \addplot[blue!30,fill=blue!20] 
        coordinates {
            (1, 561)
            (2, 849)
            (4, 1409)
            (8, 2529)
            (16, 4785)
        };
        \addlegendentry{equal};
        \addplot[blue!60,fill=blue!50] 
        coordinates {
            (1, 401)
            (2, 513)
            (4, 753)
            (8, 1233)
            (16, 2177)
        };
        \addlegendentry{cp-equal};
        \addplot[red!30,fill=red!20]
            coordinates {
            (1, 561)
            (2, 849)
            (4, 1409)
            (8, 2529)
            (16, 4785)
        };
        \addlegendentry{range};
        \addplot[red!60,fill=red!50]
            coordinates {
            (1, 401)
            (2, 513)
            (4, 753)
            (8, 1233)
            (16, 2177)
        };
        \addlegendentry{cp-range};
        \addplot[violet!30,fill=violet!20]
        coordinates {
            (1, 929)
            (2, 1569)
            (4, 2849)
            (8, 5425)
            (16, 10593)
        };
        \addlegendentry{member};
        \addplot[violet!60,fill=violet!50]
        coordinates {
            (1, 753)
            (2, 1249)
            (4, 2209)
            (8, 4129)
            (16, 7969)
        };
        \addlegendentry{cp-member};
        \nextgroupplot[title=\( \mathbf{pk^{ZK}} \)]
        \addplot[blue!30,fill=blue!20] 
        coordinates {
            (1, 75108)
            (2, 120932)
            (4, 179820)
            (8, 330356)
            (16, 631432)
        };
        \addlegendentry{equal};
        \addplot[blue!60,fill=blue!50] 
        coordinates {
            (1, 61020)
            (2, 76376)
            (4, 123468)
            (8, 184888)
            (16, 340496)
        };
        \addlegendentry{cp-equal};
        \addplot[red!30,fill=red!20] 
            coordinates {
            (1, 75148)
            (2, 121020)
            (4, 179988)
            (8, 330696)
            (16, 632112)
        };
        \addlegendentry{range};
        \addplot[red!60,fill=red!50] 
            coordinates {
            (1, 61060)
            (2, 76460)
            (4, 123640)
            (8, 185228)
            (16, 341176)
        };
        \addlegendentry{cp-range};
        \addplot[violet!30,fill=violet!20]
        coordinates {
            (1, 144156)
            (2, 259036)
            (4, 488788)
            (8, 948300)
            (16, 1867320)
        };
        \addlegendentry{member};
        \addplot[violet!60,fill=violet!50]
        coordinates {
            (1, 130068)
            (2, 230860)
            (4, 366904)
            (8, 704528)
            (16, 1379776)
        };
        \addlegendentry{cp-member};
        \nextgroupplot[title=\( \mathbf{vk^{ZK}} \)]
        \addplot[blue!30,fill=blue!20] 
        coordinates {
            (1, 8)
            (2, 12)
            (4, 16)
            (8, 28)
            (16, 48)
        };
        \addlegendentry{equal};
        \addplot[blue!60,fill=blue!50] 
        coordinates {
            (1, 8)
            (2, 12)
            (4, 16)
            (8, 28)
            (16, 48)
        };
        \addlegendentry{cp-equal};
        \addplot[red!30,fill=red!20] 
            coordinates {
            (1, 8)
            (2, 12)
            (4, 16)
            (8, 28)
            (16, 52)
        };
        \addlegendentry{range};
        \addplot[red!60,fill=red!50] 
            coordinates {
            (1, 8)
            (2, 12)
            (4, 16)
            (8, 28)
            (16, 52)
        };
        \addlegendentry{cp-range};
        \addplot[violet!30,fill=violet!20]
        coordinates {
            (1, 16)
            (2, 24)
            (4, 40)
            (8, 72)
            (16, 136)
        };
        \addlegendentry{member};
        \addplot[violet!60,fill=violet!50]
        coordinates {
            (1, 16)
            (2, 24)
            (4, 40)
            (8, 72)
            (16, 136)
        };
        \addlegendentry{cp-member};
        \end{groupplot}
    \end{tikzpicture}
     \caption{Sizes for \( Compiled \), zero-knowledge proving key \( pk^{ZK} \), and zero-knowledge verification key \( vk^{ZK} \) for equality, range, and membership proofs with respect to the number of conditions.}
    \label{fig:sizes}
\end{figure*}
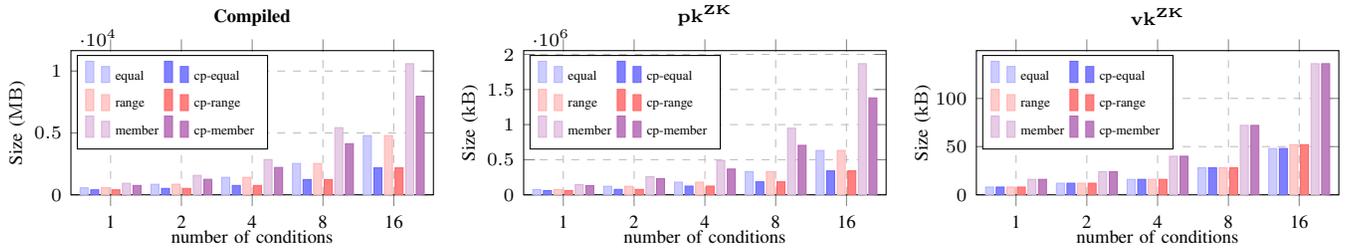

The performance experiment results for equality, range, and membership proofs are depicted in Figure~\ref{fig:performance}.
The witness phase shows a near-linear increase in time as the number of conditions grows for all proof types.
Membership proofs show the longest runtime.
Equality and range proofs have nearly the same results.
The setup phase is the most time-intensive operation for all proof types.
Membership proofs require more setup time than equality and range proofs.
Proof generation results indicate similar behavior as the number of conditions increases, aligning with the computational complexity in the witness phase.
Transaction costs for on-chain verification increase linearly with the number of conditions, reflecting the zk-SNARK-specific operations required for verification.
For example, a membership proof costs approximately \num{8114710} gas (\numusd{20.69} on the Ethereum mainnet~\cite{EthereumOrg} at the time of writing) for 16 conditions, whereas an equality proof requires around \num{2717635} gas (\numusd{6.93}), and a range proof requires around \num{3015043} gas (\numusd{7.69}).
On a Layer 2 chain the costs can be reduced (\numusd{0.04} for membership proofs, \numusd{0.01} for equality and \numusd{0.01} for range proofs on Polygon~\cite{Polygon}).
The results of measuring the sizes for \( Compiled \), \( pk^{ZK} \) and \( vk^{ZK} \) for equality, range, and membership proofs are depicted in Figure~\ref{fig:sizes}.
\sisetup{round-mode=none}
Our storage measurements show results that align with our performance findings.
The size of \( Compiled \) increases with the number of conditions, and membership proofs show the greatest storage overhead.
The size of \( pk^{ZK} \) follows a similar trend.
The \( vk^{ZK} \) shows smaller sizes for all proof types, which is needed for practical on-chain usage.

Overall, the results of equality and range proofs are almost identical. 
This is a consequence of their very similar logic, with only the condition operator being different.
By contrast, membership proofs generally involve more complex operations and inputs.

\subsubsection{Scheme comparison results}
Our results shown in Figures \ref{fig:performance} and \ref{fig:sizes} indicate, that the commit-and-prove scheme overall performs better and has less storage requirements than the baseline approach.
The only exception to this trend is observed in transaction costs and the size of \( vk^{ZK} \).
Table~\ref{tab:scheme_comparison} presents a comparison of the baseline and commit-and-prove scheme focusing on our performance and storage measurements for 16 conditions.
The result of all measurements are shown in Figures \ref{fig:performance} and \ref{fig:sizes}.
Transaction costs are omitted for readability since they remain nearly identical.
\begin{table}[htbp]
    \centering
    \caption{Scheme comparison for 16 conditions.}
    \scriptsize
    \begin{tabularx}{\columnwidth}{Xcccccc}
    \toprule
    \makecell[l]{\textbf{Proof}} & \makecell{\textbf{Witness}\\(\unit{\second})} & \makecell{\textbf{Setup}\\(\unit{\second})} & \makecell{\textbf{Prove}\\(\unit{\second})} & \makecell{\textbf{Compiled}\\(\unit{\mega\byte})} & \makecell{\( \mathbf{pk^{ZK}} \)\\(\unit{\mega\byte})} & \makecell{\( \mathbf{vk^{ZK}} \)\\(\unit{\kilo\byte})} \\
    \midrule
    equal & 19.44 & 27.84 & 26.35 & 4785 & 631 & 48 \\
    cp-equal & 8.77 & 12.12 & 10.99 & 2177 & 340 & 48 \\
    range & 19.02 & 26.74 & 25.42 & 4785 & 632 & 52 \\
    cp-range & 8.86 & 12.29 & 11.07 & 2177 & 341 & 52 \\
    member & 43.27 & 85.20 & 75.19 & 10593 & 1867 & 136 \\
    cp-member & 32.58 & 58.07 & 52.34 & 7969 & 1380 & 136 \\
    \bottomrule
    \end{tabularx}
    \label{tab:scheme_comparison}
\end{table}
The comparison indicates that the commit-and-prove scheme decreases the time required for witness generation, setup, and proving.
Storage requirements reduce for all proof types except for the size of \( vk^{ZK} \).

%% file: contents4/07_Discussion.tex
\section{Discussion}
\label{subsec:discussion}
We now revisit our design goals from Subsection~\ref{subsec:design_goals} and discuss implications on the practicality of our proposed system using the experiment results presented in Subsection~\ref{subsec:experiments}.

\subsection{Design goals}
For \textbf{decentralized trust}, VCs enable subjects to control their attributes while involving multiple trusted issuers. 
Subjects construct \( VP^{ZK} \) from these VCs, avoiding implicit trust assumptions. 
However, if an owner only accepts VCs from a single issuer, this issuer might have centralized control.
Executing policy decisions on-chain prevents single points of failure and manipulation by owners.
In a public, permissionless blockchain, this achieves decentralization of the PDP.
The same applies to the PEP and PRP, which are also executed on-chain.

Regarding \textbf{privacy preservation}, zk-SNARKs enable verification of policy results without revealing the attributes of a subject.
This supports data minimization and privacy-preserving permissioning.
Attribute inference may still occur through range proofs over small intervals or membership proofs over small sets.
Moreover, external observers may link multiple transactions to the same identity, (i) due to the pseudonymous nature of blockchain accounts, or (ii) for the on-chain signature verification in the commit-and-prove scheme.
However, subject DIDs and attributes are not revealed.
To minimize the risk of (i), the subject can avoid using the same blockchain account for multiple transactions.

Similar to decentralized trust, \textbf{transparency} of the policy enforcement, decision, and storage is achieved, since they are executed on-chain.
Although policies are referenced in the PRP on-chain, they are linked to a \( VP^{ZK} \), which may be stored off-chain in the VDR.
Therefore, transparency for policies in the PRP can only be guaranteed if the VDR is publicly accessible, which depends on the type of VDR used.
Achieving a balance between transparency and the privacy requirements of the involved parties is needed.
Therefore, only the result of the policy decision is made transparent and its underlying input values remain private.

\subsection{Practicality}
The experiment results show that more conditions result in longer runtimes and larger artifact sizes.
Membership proofs pose longer setup phases and higher memory requirements compared to equality and range proofs. 
The size of \( vk^{ZK} \) remains small, making it suitable for on-chain usage. 
The most time-intensive operation is the setup phase.
It is executed only once when creating the policy.
Transaction costs for policy decisions are considerably high and grow with more complex conditions. 
The commit-and-prove scheme performs better than the baseline approach, achieving approximately 50\% reduction for range and equality proofs and 25\% reduction for membership proofs.
This improvement is observed for all evaluated proof types, except for the transaction costs and the size of \( vk^{ZK} \).
Our current implementation employs the SHA256 hash function. 
Using a SNARK-friendly hash function like Poseidon~\cite{Poseidon} could offer improvements in runtime and storage.
For the commit-and-prove scheme, additional gas costs are required for verifying signatures on-chain.
Operating on a Layer 2 blockchain can reduce this impact.
Additionally, signature verification results can be reused for the same claim across multiple policy decisions and transactions, further reducing gas costs.
The used EdDSA signature is not natively supported by the EVM.
The Elliptic Curve Digital Signature Algorithm (ECDSA) signature scheme can be used to mitigate performance and cost issues.
Since the issuer DID is included in a claim, conflicting VCs e.g. from multiple issuers can be differentiated.
In an operational environment the design may face challenges that are not included by our specification.
Although credential revocation is currently not covered, the system supports a future integration.
Techniques for blockchain-based revocation as proposed in~\cite{BloBaAnAuSeReSmInAp} using a revocation list, or~\cite{NonDisCreOncBlocDApps} using time-dependent proofs can be used.
Regarding practicality of the demonstrated use case, the presented system can be used as a foundation to support advanced KYC/AML checks.

%% file: contents4/08_Conclusion.tex
\section{Conclusion}
\label{sec:conclusion}
In this paper, we introduced a general model for privacy-preserving on-chain permissioning that combines SSI and ZKPs.
It addresses the challenge of permissioning in decentralized applications operating on public blockchains, where KYC/AML obligations must be satisfied without compromising transparency, privacy or the decentralized nature of blockchains.
Our system integrates concepts such as DIDs and VCs, facilitating a user-centric and decentralized identity management.
We described the components and interactions of our system, outlining the transition from an SSI-based to a privacy-preserving system design for on-chain permissioning, incorporating zk-SNARKs to support non-disclosing policy decisions.
A practical instantiation of our general model demonstrates applicability to KYC/AML compliance in a DeFi scenario with support for equality, range, membership, and time-dependent proofs.
We provided guidelines for practical realization, including a commit-and-prove scheme that verifies credential authenticity outside the ZKP circuit.
Our evaluation proved feasibility by providing an implementation to perform zkSNARKs-based policy decisions using ZoKrates~\cite{ZoKrates} on an EVM-based blockchain and gives first insights into the performance of our proposal by conducting an initial set of experiments.
The commit-and-prove scheme reduces performance and storage requirements by 50\% for range and equality proofs, and 25\% for membership proofs regarding zk-SNARK specific operations.